\documentclass{aastex6}
\usepackage{float}
\usepackage[normalem]{ulem}

\begin{document}

\title{Evidence for a systematic offset of $-$80~micro-arcseconds in the {\it Gaia\/} DR2 parallaxes}
\author{Keivan G.\ Stassun\altaffilmark{1,2} and Guillermo Torres\altaffilmark{3}}
\altaffiltext{1}{Vanderbilt University, Department of Physics \& Astronomy, 6301 Stevenson Center Ln., Nashville, TN  37235, USA%; {\tt keivan.stassun@vanderbilt.edu}
}
\altaffiltext{2}{Fisk University, Department of Physics, 1000 17th Ave. N., Nashville, TN  37208, USA}
\altaffiltext{3}{Harvard-Smithsonian Center for Astrophysics, 60 Garden St., Cambridge, MA 02138, USA}

\begin{abstract}
%We previously used the sample of eclipsing binaries with accurate, empirical distances from \citet{Stassun:2016} to test the parallaxes reported in the {\it Gaia\/} first data release, finding an average offset of $-$0.25$\pm$0.05~mas in the sense of the {\it Gaia\/} parallaxes being too small (i.e., the distances too long). 
We reprise the analysis of \citet{Stassun:2016}, comparing the parallaxes of the eclipsing binaries reported in that paper to the parallaxes newly reported in the {\it Gaia\/} second data release (DR2). 
We find evidence for a systematic offset
%, at 98\% confidence, 
of $-$82$\pm$33~$\mu$as, in the sense of the {\it Gaia\/} parallaxes being too small, for brightnesses $(G \lesssim 12)$ and for distances (0.03--3~kpc) in the ranges spanned by the eclipsing binary sample. 
The offset does not appear to depend strongly on distance within this range, though there is marginal evidence that the offset increases (becomes slightly more negative) for distances $\gtrsim$1~kpc, up to the 3~kpc distances probed by the test sample. 
The offset reported here is consistent with the expectation that global systematics in the {\it Gaia\/} DR2 parallaxes are below 100~$\mu$as.
\end{abstract}

\section{Introduction}\label{sec:intro}
The trigonometric parallaxes for $\sim$10$^9$ stars from the {\it Gaia\/} mission promise to revolutionize many areas of stellar astrophysics. 
For example, 
%with eventual expected precision in the parallax of $\approx$20~$\mu$as for bright exoplanet host stars, 
%it should be possible to determine the stellar and planet radii and masses directly and empirically with accuracies of 3--5\% \citep[see, e.g.,][]{Stassun:2017}. As another example, 
it will be possible to determine accurate radii and masses for tens of thousands of bright stars observed by the Transiting Exoplanet Survey Satellite \citep[TESS;][]{Ricker:2011}, by combining the {\it Gaia\/} parallaxes with the granulation ``flicker" \citep{Bastien:2016} in the TESS light curves \citep{Stassun:2017,Stassun:2018}.

%It is essential to assess the on-sky delivered performance of these parallaxes from {\it Gaia\/}, especially the presence of any unexpected biases. This is particularly important in light of the experience from {\it Hipparcos\/}, which suffered a significant bias in at least the case of the Pleiades cluster \citep[e.g.,][]{Pinsonneault:1998}. 
%Such a check requires a set of benchmark stars whose parallaxes are determined independent of the {\it Gaia\/} parallaxes, and preferably independent of parallaxes altogether. 

In order to perform an independent assessment of the {\it Gaia\/} parallaxes,
\citet{Stassun:2016} assembled a sample of 158 eclipsing binary stars (EBs) whose radii and effective temperatures are known empirically and precisely, such that their bolometric luminosities are determined to high precision (via the Stefan-Boltzmann relation) and therefore independent of assumed distance. 
\citet{Stassun:2016} also measured the bolometric fluxes for these EBs which, together with the precisely known bolometric luminosities, yielded the EB distances;
%
%The precision of the parallaxes for this EB sample was predicted by \citet{Stassun:2016} to be 
the precision on the predicted parallaxes is $\approx$190~$\mu$as on average. 
%This is a factor of $\sim$1.5 better than the median precision of 320~$\mu$as for {\it Gaia\/} DR1 \citep{Gaia:2016}. It is even somewhat superior to the expected {\it Gaia\/} DR1 precision floor of 240~$\mu$as. %
While this precision is poorer than that expected from the {\it Gaia\/} second data release (DR2), for which the global systematics are expected to be below 100~$\mu$as \citep{Gaia:2018}, the EB sample is large enough that it should be possible in principle to assess average systematics down to $\sim 190/\sqrt{158}\sim 15$~$\mu$as.
%These EB parallaxes can therefore readily serve as distance benchmarks for the trigonometric parallaxes reported by {\it Gaia} DR2, and in particular can be used to assess the presence of any systematics. 

In \citet{StassunTorres:2016} we reported an initial assessment of the {\it Gaia\/} first data release (DR1) parallaxes, finding a significant average offset of $-0.25\pm 0.05$~mas, in the sense that the {\it Gaia\/} DR1 parallaxes were too small (i.e., the {\it Gaia\/} distances too long), at least within the range of distances probed by the overlap of the EB and {\it Gaia\/} DR1 samples ($\lesssim$1~kpc). That finding, which was consistent with the expected systematic error floor of 0.3~mas for {\it Gaia\/} DR1 \citep{Gaia:2016}, was corroborated by other authors on the basis of ground-based parallaxes of nearby M~dwarfs \citep{Jao:2016} and asteroseismic stellar radii in the Kepler field \citep{Huber:2017}.
%, as well the small subset of nearby Cepheids in the study of \citet{Casertano:2017}. 
The study of \citet{Huber:2017} found a somewhat smaller offset and also discussed potential biases in the stellar luminosities arising from systematics in the effective temperature scale. Another analysis on the basis of stellar radii from granulation ``flicker" \citep{Bastien:2013,Bastien:2016,Corsaro:2014} again corroborated the \citet{StassunTorres:2016} offset \citep{Stassun:2018}. 

In this paper, we report the results of testing the {\it Gaia\/} DR2 parallaxes against the same \citet{Stassun:2016} EB benchmark sample as we used in \citet{StassunTorres:2016}. 
Our intent is to provide an additional independent validation to those considered by the {\it Gaia\/} Mission team \citep{Arenou:2018}, which finds evidence for a global parallax offset ranging from about $-30$~$\mu$as to about $-50$~$\mu$as for most of the benchmark samples they considered.
Section~\ref{sec:data} summarizes the EB and {\it Gaia\/} data used. Section~\ref{sec:results} presents the key result of a systematic offset in the {\it Gaia\/} parallaxes relative to the EB sample. Section~\ref{sec:disc} considers potential trends in the parallax offset with other parameters. Section~\ref{sec:summary} concludes with a summary of our conclusions.

\section{Data}\label{sec:data}
We adopted the predicted parallaxes for the 158 EBs included in the study of \citet{Stassun:2016}. Of these, 151 have parallaxes available in {\it Gaia\/} DR2.
%\footnote{Accessed on 14 September 2016.}. 
We excluded from our analysis any parallaxes identified in the {\it Gaia\/} release as potentially problematic \citep[{\tt ASTROMETRIC\_EXCESS\_NOISE\_SIG} flag greater than 2; see][]{Lindegren:2012}, leaving 89 EBs, none of which were flagged as potentially problematic by \citet{Stassun:2016}. 
Thus our primary study sample is 89 EBs with good parallaxes from both the EB analysis and from {\it Gaia} DR2. 
These EBs are all relatively nearby, with parallaxes in the range $\pi \approx$ 0.3--30~mas.
%The EBs and their relevant data are provided in Table~\ref{tab:data}.
Finally, all 89 EBs have {\it Gaia\/} parallax errors better than 15\%, and thus the choice of prior on the parallax should not be important for inferring distances \citep[e.g.,][]{Bailer-Jones:2016}.

\section{Results}\label{sec:results}

\begin{figure}[!ht]
\centering
\includegraphics[width=0.7\linewidth,trim=0 80 10 60,clip]{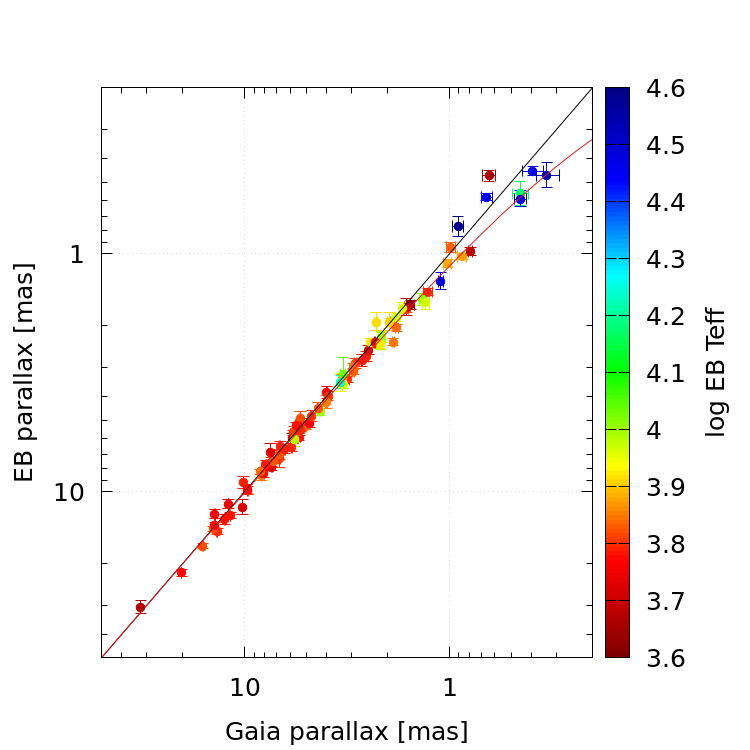}
\includegraphics[width=0.71\linewidth,trim=13 0 10 70,clip]{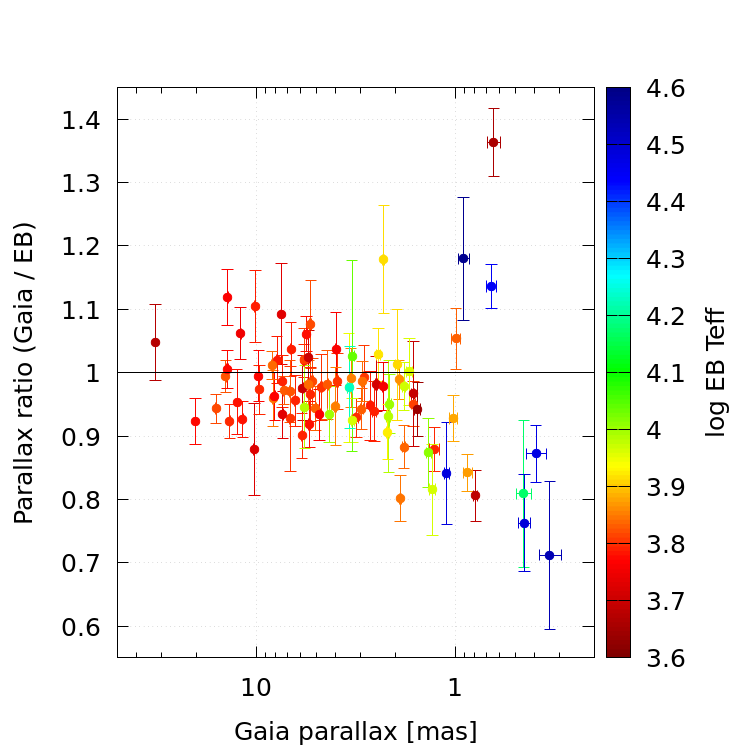}
\caption{{\it (Top:)} Direct comparison of predicted parallaxes from the eclipsing binary sample of \citet{Stassun:2016} versus the parallaxes from the {\it Gaia\/} second data release. The one-to-one line is shown in black and a least-squares linear best fit is shown in red. Color represents $T_{\rm eff}$ for the EBs, a proxy for system color.
{\it (Bottom:)} Same as top, except showing ratio of $\pi_{\rm EB}$ to $\pi_{\rm Gaia}$.}
\label{fig:pxvspx} 
\end{figure}

Figure~\ref{fig:pxvspx}a shows the direct comparison of the EB parallax predictions from \citet{Stassun:2016} versus the {\it Gaia\/} DR2 parallaxes for the study sample. The least-squares linear best fit, weighted by the measurement uncertainties in both quantities \citep{Press:1992}, is $\pi_{\rm EB} = 0.062 (\pm 0.047) + 1.015 (\pm 0.007) \times \pi_{\rm Gaia}$. 
%While this indicates a good 1-to-1 agreement to first order, the coefficient of 1.03$\pm$0.01 could be interpreted as a modest global difference of {\it scale} in the {\it Gaia\/} parallaxes relative to the EB parallaxes. However, considering all of the available evidence instead suggests a small {\it offset} in the {\it Gaia\/} parallaxes 
%\sout{among stars with relatively large parallaxes}, 
%as we now discuss. 
To first order, the agreement between the {\it Gaia\/} DR2 and EB parallaxes is excellent. However, the linear fit coefficients do indicate that the EB parallaxes may be larger on average than the {\it Gaia\/} parallaxes, and moreover the deviation from 1-to-1 agreement in Fig.~\ref{fig:pxvspx}a hints at the relative difference becoming somewhat larger at smaller parallaxes (see also Fig.~\ref{fig:pxvspx}b).

\begin{figure}[!ht]
    \centering
    \includegraphics[width=0.68\linewidth,trim=10 0 10 75,clip]{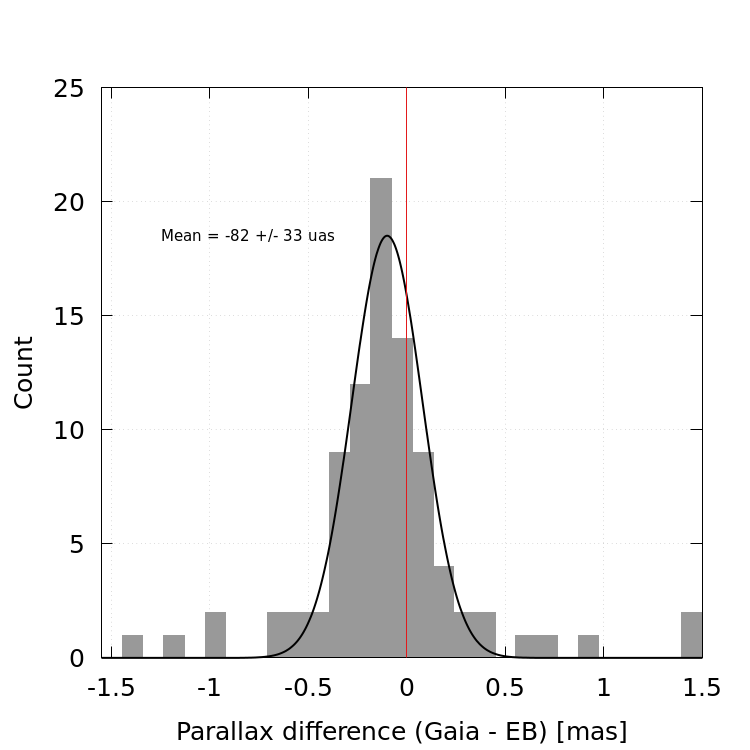}
    \caption{Distribution of $\Delta\pi$ ({\it Gaia\/}$-$EB). EBs with known tertiary companions are represented in blue. We find a weighted mean offset of $\Delta\pi = -82 \pm 33$~$\mu$as. Also shown is a best-fit Gaussian with $\sigma^2 = \sigma_{\rm Gaia}^2 + \sigma_{\rm EB}^2$, representing the quadrature sum of the typical random uncertainties from the {\it Gaia\/} and EB parallaxes; see Sec.~\ref{sec:intro}.}
    \label{fig:delta_plx_hist}
\end{figure}

Figure~\ref{fig:delta_plx_hist} presents the overall distribution of parallax differences in the sense of $\pi_{\rm Gaia} - \pi_{\rm EB}$. 
The distribution appears roughly symmetric and normally distributed, with perhaps a sharper peak and more extended
wings than a Gaussian, and there is a clear offset relative to zero. 
The weighted mean offset is $\Delta\pi = -82 \pm 33$~$\mu$as, where the quoted error is the uncertainty of the mean for 89 measurements. 
Note that weighted mean offset for the full sample is similar, $\Delta\pi = -107 \pm 28$~$\mu$as, though to be clear we do not quote this as our main result due to the {\it Gaia\/} parallaxes being reported as unreliable for many of these (see Sec.~\ref{sec:data}). 

%\citet{Stassun:2016} noted that a number of the EBs used in that study are known triple or quadruple systems. In general these companions contribute very little to the total system light, and \citet{Stassun:2016} found no evidence for significant systematics in their predicted parallaxes. Nonetheless, the offset that we find above for the {\it Gaia\/} parallaxes is small and could potentially have gone unnoticed in \citet{Stassun:2016}. Indeed, the effect of additional light contribution to the EBs by companions would be in the sense of making the EB stars appear brighter, therefore inferred to be closer, and in turn the {\it Gaia\/} distances interpreted as too long (parallax too small). 

%In {\it Gaia\/} DR1, 12 of our EBs have known companions (see Table~\ref{tab:data}; one additional EB with a known companion is already excluded by the cuts discussed in Sec.~\ref{sec:data}). These EBs are indicated in blue in Fig.~\ref{fig:delta_plx_hist}, which shows the two largest outliers to be triples. Excluding all of the triples results in a parallax offset of 
%$\Delta\pi = -0.233 \pm 0.046$ mas, consistent with that determined for the full sample, though slightly smaller. 

%Overall, from the EB sample, a systematic offset in the {\it Gaia\/} DR1 parallaxes of $-0.233$ to $-0.264$~mas is indicated. 
%For simplicity, we adopt the rounded value between these estimates of $-0.25\pm 0.05$~mas.

\section{Discussion}\label{sec:disc}

The official {\it Gaia\/} DR2 documentation states\footnote{\url{http://www.cosmos.esa.int/web/gaia/dr2}}: 
``Parallax systematics exist depending on celestial position, magnitude, and colour, and are estimated to be below 0.1~mas. There is a significant average parallax zero-point of about $-30$~$\mu$as."
Our finding of a mean parallax offset of $\Delta\pi\approx -82 \pm 33$~$\mu$as (Sec.~\ref{sec:results}; Fig.~\ref{fig:delta_plx_hist}) is consistent with this expectation of a systematic error floor below 0.1~mas.

In principle this offset could be due to systematics in one or more of the EB parameters from which the EB distances are determined. If so, one might especially suspect the EB $T_{\rm eff}$ values: unlike the stellar radii, for example, which are determined from simple geometry, the $T_{\rm eff}$ values are determined from spectral analysis and/or spectral typing and/or color relations. 
The slope of the fitted relation in Fig.~\ref{fig:pxvspx} would imply an error in the EB distance scale of $\sim$1.5\%, which in turn would require a systematic error in $T_{\rm eff}$ of $\sim$0.75\% (because $d \sim L_{\rm bol}^{1/2} \sim T_{\rm eff}^2$) or $\sim$50~K given the typical $T_{\rm eff}$ of the EB sample. 
The sense of the offset is that the EBs would have to be systematically too cool. 

This possibility was also considered in our previous study of the DR1 parallax offset \citep{StassunTorres:2016} and was discounted as unlikely for multiple reasons, 
%However, we do not consider this to be a likely possibility, for multiple reasons. 
%For example, \citet{Stassun:2016} found no evidence for a systematic offset of the EB parallaxes relative to 
including comparison to the {\it Hipparcos\/} parallaxes 
%which, even at the somewhat poorer precision of $\sim$1~mas, should have been apparent. 
%\citet{Lindegren:2016} compare the {\it Gaia\/} DR1 parallaxes against $\sim$87,000 {\it Hipparcos\/} stars in common, finding a statistically significant average offset just under $-0.1$~mas, smaller than, but in the same sense as the offset we find among our EB sample.
%In addition, because 
and the fact that the various EBs in our study sample have had their $T_{\rm eff}$ determined by different methodologies and different calibrations, 
%adopted by the various authors, it is 
making it very unlikely that any individual biases should produce a net systematic offset in a sample of 89 EBs spanning a large range of $T_{\rm eff}$. 
%While it is true that typical systematics among various $T_{\rm eff}$ scales can be of order 50--100~K \citep[see, e.g.,][]{Casagrande:2011,Heiter:2015} and that some of the methods are not entirely independent, others are, and we consider it highly unlikely that the whole EB sample would have their $T_{\rm eff}$ systematically offset by that magnitude.
%
Finally, we have directly examined the degree to which $\Delta\pi$ correlates with $T_{\rm eff}$ in the EB sample (Fig.~\ref{fig:pxvspx}).
It is true that the hotter stars tend to have the largest fractional $\Delta\pi$, however this is most likely a consequence of the fact that the hotter (more luminous) EBs tend to be at larger distances and that a constant $\Delta\pi$ is largest fractionally at small $\pi$.
Moreover, fractionally large $\Delta\pi$ also occur among cool EBs in Fig.~\ref{fig:pxvspx}. 
%Indeed, a Kendall's $\tau$ non-parametric correlation test between $\Delta\pi$ and $T_{\rm eff}$ is not significant (67\% probability that the null hypothesis---that $\Delta\pi$ and $T_{\rm eff}$ are uncorrelated---can be rejected). 
%(We checked that the parallax {\it ratio} versus $T_{\rm eff}$ is also not significantly correlated.)
%Incidentally, this also suggests little to no dependence of $\Delta\pi$ with color, since $T_{\rm eff}$ can be taken as a proxy for color. 

%\begin{figure}[!ht]
%    \centering
%    \includegraphics[width=0.9\linewidth,trim=10 0 10 60,clip]{delta_vs_teff.png}
%    \caption{Examining potential correlation between fractional $\Delta\pi$ and EB $\log T_{\rm eff}$. 
%    The black curve represents the relationship between $\Delta\pi/\pi$ and $\pi$ if $\Delta\pi$ is a constant as found in Sec.~\ref{sec:results}. }
%    \label{fig:delta_vs_eb_plx}
%\end{figure}

We also examined possible correlations of $\Delta\pi$ with stellar brightness, color, and position on sky. No statistically significant patterns were found, however the sample here may not be large enough to detect low-level, multivariate dependencies on these parameters; the {\it Gaia\/} DR2 documentation suggests that such low-level dependencies are likely to be present \citep{Arenou:2018}.
Indeed, the EBs in our sample are all relatively bright, $G\lesssim 12$, and thus may not adequately probe magnitude-dependent effects. 

Finally, we consider an alternate interpretation to a constant systematic error (zero-point offset). We can reframe the linear fit shown in Fig.~\ref{fig:pxvspx}a in terms of both a zero-point offset and an error in scale.
The fitted relation between $\pi_{\rm EB}$ and $\pi_{\rm Gaia}$ in Sec.~\ref{sec:results} can be rewritten as $\pi_{\rm Gaia}-\pi_{\rm EB} = -0.061-0.015\times \pi_{\rm EB}$. Note that for sufficiently distant stars, $\pi_{\rm EB}$ is necessarily more precise than $\pi_{\rm Gaia}$, since the relative precision on EB-based parallaxes do not depend on distance \citep{Stassun:2016}. Thus, the intercept of the linear fit ($-0.061$~mas) becomes the estimator of the zero-point offset in the $\pi_{\rm Gaia}$ parallaxes. In that case, the fitted slope coefficient ($-0.015$) would correspond to an error in {\it scale}, which could in principle be produced by systematics in the EB parameters (e.g.\ $T_{\rm eff}$ systematically too small; see above) and/or by magnitude- or color-dependent systematics in {\it Gaia}, if magnitude or color are correlated with distance in the sample. Indeed, as noted above (see Fig.~\ref{fig:pxvspx}), the EB sample does indeed possess some correlation between color and distance, due simply to the fact that the hottest EBs have the highest luminosities and therefore can be observed at larger distances. In any event, the zero-point offset in this case---in which there is both an offset error and a scale error---would be $-61\pm 46$~$\mu$as. This offset is somewhat smaller than (and less statistically significant), but consistent with, the value of $-82\pm 33$~$\mu$as that we found above (Sec.~\ref{sec:results}) as a global mean offset. 

In comparison to our findings reported here, 
a recent analysis of the {\it Gaia\/} DR2 parallaxes relative to a sample of benchmark Cepheids finds evidence for a global offset of $-46\pm 13$~$\mu$as \citep{Riess:2018}.
Similarly, a concurrent independent analysis by \citet{Zinn:2018} finds a nearly identical systematic offset of $-52.8 \pm 2.4$~$\mu$as (statistical) $\pm 1$~$\mu$as (systematic) using a large sample of more than 3000 stars with asteroseismically determined radii. These results are perhaps in better agreement with the offset of $-61\pm 46$~$\mu$as that we find above when considering both zero-point and scale terms; indeed, the \citet{Zinn:2018} preliminary analysis suggests that the offset may increase somewhat at larger distances (smaller parallaxes), similar to the hint of such a trend that we observe among the handful of EBs at distances beyond $\sim$1~kpc (Fig.~\ref{fig:pxvspx}), which again may suggest a small error in scale in addition to a zero-point systematic error. Alternatively, we note that in the \citet{Zinn:2018} sample, the brighter stars with $G<11$ (more similar in brightness to our EB sample) appear to exhibit a larger systematic offset of $\approx -80$~$\mu$as, similar to the global offset that we find.
Finally, an analysis by \citet{Kounkel:2018} of 55 young stars with Very Large Baseline Array parallaxes finds an offset of $-74\pm 34$~$\mu$as, with mild evidence for a scale error term (at 1$\sigma$ confidence).
In any case, our finding of a simple global offset of $\Delta\pi = -82\pm 33$~$\mu$as is fully consistent with these other reports and is, we believe, a reasonably good estimate of the offset for relatively bright ($G\lesssim 12$) and nearby ($d\lesssim 1$~kpc) stars as represented by our EB sample.

\section{Summary and Conclusions}\label{sec:summary}

Here we present evidence of a small but systematic offset in the average
zero-point of the parallax measurements recently released by the
{\it Gaia\/} Mission of about $-82 \pm 33$~$\mu$as, in the sense that
the {\it Gaia\/} values are too small. 
To apply the correction, this (negative) offset must be {\it subtracted} from the reported {\it Gaia\/} DR2 parallaxes. 
Because the benchmark sample that we use is mainly within $\sim$1~kpc, we can only confirm that the offset is statistically valid for relatively large parallaxes, $\pi \gtrsim 1$~mas.
In addition, the sample used here is relatively bright, with $G\lesssim 12$.

The reference for this determination is a set of 89 independently inferred parallaxes from a benchmark sample of well-studied eclipsing binaries with a wide range of brightnesses and distributed over the entire sky. 
This paper presents evidence of a {\it difference} between the {\it Gaia\/} and EB parallaxes, which we have interpreted here as a systematic error in {\it Gaia\/} after discussing the alternative. In particular, we have considered the possibility of a systematic offset in the EB effective temperature scale 
%as the most plausible, but very unlikely, possibility.
as a possible, but unlikely alternative explanation. 
%Other authors \citep[][]{Riess:2018,Zinn:2018} find a nearly identical result using large samples of Cepheids and of stars with asteroseismically determined radii, respectively. 
Other authors who have performed concurrent, independent analyses with different benchmark samples \citep{Riess:2018,Zinn:2018,Kounkel:2018} report similar offsets in the range from $-46$~$\mu$as to $-74$~$\mu$as.
We have also discussed, and cannot rule out, the possibility that there exists a small error in scale as well as an error in zero-point; in that case the zero-point offset becomes $-61\pm 46$~$\mu$as for the EB sample studied here. 
%In any event, our finding of an offset in the range $-60$~$\mu$as to $-80$~$\mu$as is similar to findings by other authors who have performed concurrent, independent analyses with different benchmark samples \citep{Riess:2018,Zinn:2018,Kounkel:2018}, whose reported offsets range from $-46$~$\mu$as to $-74$~$\mu$as.

Finally, the parallax offset independently reported here is a testament to the quality of the vetting of the {\it Gaia\/} Mission. 
%Just as the $-0.25$~mas offset we reported in \citet{StassunTorres:2016} was consistent with the expected 0.3~mas systematic error floor for the {\it Gaia\/} DR1 parallaxes \citep{Gaia:2016}, 
The $-80$~$\mu$as offset we find here is perfectly consistent with the 100~$\mu$as systematic error floor reported by the {\it Gaia\/} Mission for DR2 \citep{Gaia:2018,Arenou:2018}.

\acknowledgments
This work has made use of the Filtergraph data visualization service at {\tt \url{filtergraph.vanderbilt.edu}} \citep{Burger:2013}. K.G.S.\ acknowledges partial support from NSF PAARE grant AST-1358862. G.T.\ acknowledges partial support for this work from NSF grant AST-1509375.
The authors are grateful to J.~Zinn for sharing their results in advance of publication. 
We are grateful to the referee for critiques and suggestions that improved the manuscript.
This work has made use of data from the European Space Agency (ESA) mission {\it Gaia\/} (http://www.cosmos.esa.int/gaia), processed by the {\it Gaia\/} Data Processing and Analysis Consortium (DPAC, http://www.cosmos.esa.int/web/gaia/dpac/consortium). Funding for the DPAC has been provided by national institutions, in particular the institutions participating in the {\it Gaia\/} Multilateral Agreement.

\end{document}